\documentclass[epsfig]{emulateapj} 

\received{} 
\revised{} 
\accepted{}

\newcommand{\msun}{M_{\odot}}

\shorttitle{Transtition from Pop III to Pop II} \shortauthors{Fang \&
  Cen}

\begin{document}

\title{Transition From Population III to Population II Stars}

\author{Taotao Fang\altaffilmark{1} and Renyue Cen\altaffilmark{2}}

\begin{abstract}

The transition from Population III to Population II stars is
determined by the presence of a sufficient amount of metals, in
particular, oxygen and carbon.  The vastly different yields of these
relevant metals between different initial stellar mass functions would then
cause such a transition to occur at different times.  We show that the
transition from Pop III to Pop II stars is likely to occur before the
universe can be reionized, if the IMF is entirely very
massive stars ($M\ge 140\,\msun$). A factor of about $10$ more
ionizing photons would be produced in the case with normal top-heavy IMF
(e.g., $M\sim 10-100\,\msun$), when such a transition occurs.
Thus, a high Thomson optical depth ($\tau_e \ge 0.11-0.14$) may be
indication that the Population III stars possess a more conventional
top-heavy IMF.

\end{abstract}


\altaffiltext{1}{Department of Astronomy, University of California,
  Berkeley, CA 94720; {\sl Chandra} Fellow}

\altaffiltext{2}{Department of Astrophysical Sciences, Princeton
  University, Peyton Hall, Ivy Lane, Princeton, NJ 08544}

\section{Introduction}

Recent observations of the high redshift ($z > 6$) quasar spectra from the
Sloan Digital Sky Survey (SDSS) and the cosmic microwave
background fluctuations from the Wilkinson Microwave Anisotropy Probe ({\sl
WMAP}) combine to paint a complicated reionization picture
(\citealp{fan01,kog03}).  While it seems relatively secure to claim that the
reionization process has begun at $z\ge 10$ and ends at
$z\sim 6$ (e.g., \citealp{fan01,bec01,bar02,cen02}), exactly when it starts
and how it evolves are yet unclear, although the overall
picture is consistent with a physically motivated double reionization model
(\citealp{cen03,wyi03}) that was proposed before WMAP released its
reionization measurement. The conventional wisdom is that stars are
primarily responsible for producing most of the ionizing photons.
Unavoidably, there is a transition epoch from metal-free Population III (Pop
III) to metal-poor Population II (Pop II) stars at some high redshift (see,
e.g., \citealp{omu00,bro01,sch02,mac03,sch03,bro03}).

It is thought that Pop III stars may be much more massive than Pop II stars.
 This expectation is based on the physics of metal
cooling. Lack of metals in the gas at early times results in an
absolute floor temperature of gas at $\sim 100$K, whereas a small but
significant amount of metals would enable gas to cool down to $\sim 10$K.
Consequently, Pop III stars are expected to be much more
massive than Pop II stars
(\citealp{car84,lar98,abe00,her01,bro01,bkl01,nak01,bro02,omu03,mac03}).
However, how massive Pop III stars are remains unclear.  While
simulations have suggested that Pop III stars may be more massive than
$100\,\msun$ (``vary massive star'', VMS; \citealp{abe00,bro01}),
\citet{tan04} argue that taking into account feedback processes would likely
limit the mass of the Pop III stars to the range
$30-100\,\msun$.  Observationally, the VMS picture is advocated by
\citet{ohn01} and \citet{qia02}, based on the argument that the metal yield
patterns from pair-instability supernova (PISN) explosion of VMS progenitors
\citep{heg02} are consistent with observations (see also Wasserburg \&
Qian~2000; 2001).  On the other hand, \citet{tum04}
argue that the general pattern in metal-poor halo stars, especially the
Fe-peak and $r$-process elements, favors the yield pattern of Type II
supernovae (SNII) with an initial mass function (IMF) above 10
$\msun$ and without VMS (see also \citealp{dai04}).  Other arguments based
on observations such as metallicities of the intergalactic (IGM) at $z \sim
3 - 4$ in the Ly$\alpha$ forest \citep{ven03} and cosmic star formation
history \citep{dai04} also favor a SNII type
IMF. \citet{ume03} and \citet{umed03} argue that the observed metal
abundance pattern are better matched by core-collapsed supernova with
$M_{\star} = 10-50\,\msun$ through pair-instability supernova
explosion \citep{heg02}.

It thus seems beneficial to explore possible differences between
different IMFs.  In this {\it Letter}, we investigate the issue of the
transition from Pop III to Pop II.  Specifically, following
\citet{bro03}, we note that the transition from Pop III to Pop II is
dictated by efficiency of cooling by a limited number of species, in
particular, C and O.  Thus, it is the amount of C and O, not
necessarily the total amount of ``metals", that determines the
transition.  While the ionizing photon production efficiency turns out to be
quite comparable in both cases, either with a VMS IMF or a more normal
top-heavy IMF (e.g., \citealp{tum04}), they have very different metal
patterns in the exploded final products.  In PISN case the
supernova ejecta is enriched by $\alpha$-elements, whereas the major
products of SNII are hydrogen and helium with a small amount of heavy
elements (see, e.g., \citealp{woo95,heg02}).  Consequently, the
transition from Pop III to Pop II stars should occur at different
times.  We show that a normal top-heavy IMF is preferred to VMS and probably
required if Pop III stars were to reionize the universe at high enough
redshift.

\section{Transition from Pop III to Pop II}

\begin{figure}
\begin{center}
\resizebox{3.5in}{!}{\includegraphics{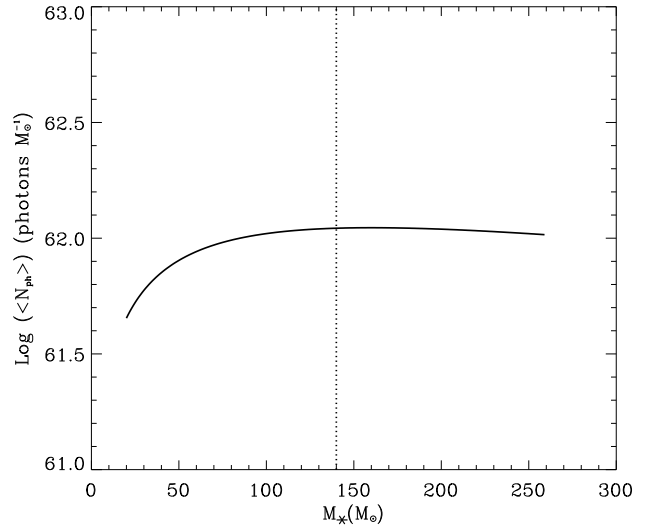}}
\end{center}
\caption{Ionization photons produced during the lifetime of a
  zero-metallicity Pop III star as a function of its initial mass
  $M_{\star}$, adopted from \citet{scha02}. The dotted vertical line
indicates $M_{\star}=140\,\msun$. Evidently, the produced ionization
photons is nearly a constant for both PISN and SNII cases.}
\label{fig:photon}
\vskip7pt
\end{figure}

We first define $f_{\star}$ as the baryon fraction that has been
processed through Pop III stars, $f_{\star} \equiv M_{III}/M_b$, where $M_b$
and $M_{III}$ are total baryonic mass and the baryonic mass in Pop III
stars, respectively. We also define the total ionization
photons ($h\nu \ge 13.6$eV) produced per unit mass of Pop III stars as
$<N_{ph}>\,\rm photons\,M_{\odot}^{-1}$. For completeness we show in
Figure~\ref{fig:photon} the time-integrated ionization photons per solar
mass emitted during the lifetime of a Pop III star vs. its
initial mass $M_{\star}$, adopted from \citet{scha02}. The vertical dotted
line at $M_{\star}=140\,\msun$ is to distinguish between PISN and SNII.
Clearly, for most of the mass range ($80\lesssim M_{\star} < 260\,\msun$),
the ionizing photons produced during the lifetime of the star is $\sim
10^{62}\,\rm photons\,M^{-1}_{\odot}$. This
lifetime-integrated number of ionizing photons varies only about a factor of
less than 2 when the initial mass is $\sim 20\,\msun$. It is because the
total ionizing photon production of a short-lifetime VMS is boosted by its
higher production rate. However, we should keep in mind that photon
production rate could be important if the typical recombination time scale
is shorter than the lifetime of Pop III
stars. The typical lifetimes of a 200 and 25 $M_{\odot}$ Pop III stars are
$\sim 2.2\times 10^6$ and $6.5\times10^6$ yrs \citep{scha02}. The
recombination time scale is roughly $t_{rec} \sim 10^7 (C_{H
II}/10)^{-1}$ yrs at redshift $z \sim 17$, where $C_{H II}$ is the clumping
factor of \ion{H}{2} regions \citep{ric04}. While in general $t_{rec}$ is
longer than the lifetime of Pop III stars, at some
overdense regions when $C_{H II} > 20$ the lifetime of a small mass star
could be longer than $t_{rec}$, and in this case more than 1
ionizing photon are necessary to ionize the entire IGM. Having this caution
in mind, in the following we assume that all the Pop III stars produce about
the same amount of ionizing photons in their lifetime and roughly 1 ionizing
photon per atom is needed to reionize the
universe. Using relevant numbers and assuming a mean atom weight of $\mu =
0.76$, we find that the number of the ionizing photons per
hydrogen atom produced, denoted as $N_{ion}$, is
\begin{equation}
N_{ion} = 11
\,\left(\frac{\mu}{0.76}\right)^{-1}\left(\frac{f_{\star}}{10^{-4}}\right)\left(\frac{<N_{ph}>}{10^{62}\,\rm
photons\,M^{-1}_{\odot}}\right).
\label{eq:Nion}
\end{equation}

Let us define $y_i$ as the yield of element $i$, $y_i\equiv
M_i/M_{\star}$, where $M_i$ is the mass of element $i$ produced by the
initial stellar mass $M_{\star}$.  Assuming metals produced by
supernova explosion can be effectively transported out of the host halos and
uniformly pollute the intergalactic medium (IGM), then the IGM metallicity
of element $i$, $Z_{IGM}^i=f_{\star}y_i$, where
$f_{\star}$ is the baryon fraction defined above. Substituting
$f_{\star}$ from Eq.~\ref{eq:Nion}, we can then relate the ionization photon
per H atom, $N_{ion}$, to the  IGM metallicity of element $i$, $Z_{IGM}^i$:
\begin{equation}
Z_{IGM}^i = \frac{\mu y_i}{<N_{ph}>m_H} N_{ion}.
\label{eq:ZIGM}
\end{equation}

Zero-metallicity Pop III stars with different initial masses have
distinctive pattern of final metal productions (see, e.g.,
\citealp{heg02}). For stars with initial masses $10 \lesssim M_{\star}
\lesssim 40\,\msun$, neutron stars are formed through type II
supernova explosion. The majority of the ejecta are hydrogen and
helium, with small amount of heavy metals. Between $40 \lesssim
M_{\star} \lesssim 100\,\msun$, black holes form through direct
collapse of the star. At $M_{\star} \sim 100\,\msun$, the
electron-positron pair instability starts to dominate, but if
$M_{\star} \lesssim 140\,\msun$, an iron core will still be formed and
eventually collapse to a black hole without any metal
productions. When $140 \lesssim M_{\star} \lesssim 260\,\msun$, the
pulsation induced by the pair instability is so violent that one
single pulse disrupts the entire star and eject abundant
$\alpha$-elements. At above $260\,\msun$ again black holes will
form. An upper limit of $500\,\msun$ of a Pop III star that can be formed
from accretion into primordial protostar was recently found by \citet{bro04}
via three-dimensional numerical simulation. It is
important to note that essentially all metals are produced for mass ranges
of 10 -- 40 $\msun$ and/or 140 -- 260 $\msun$. Pop III stars within these
two mass ranges produce distinctive pattern of metals, which will have great
impact on the important metal cooling agents, namely, C and O.

We study four types of IMFs: two types for PISN and two types for Type II
SN. An IMF is defined as $\Psi(M) \equiv \Psi_0M^{-\Gamma}$ with $\Psi_0$
\begin{equation}
\int_{m_1}^{m_2} \Psi_0 M^{-\Gamma} dM = 1.
\end{equation} The four types are: (1) PISN model 1, with $140 \lesssim
M_{\star} \lesssim 260\,\msun$; (2) PISN model 2, $100 \lesssim
M_{\star} \lesssim 260\,\msun$; (3) Type II Supernova model 1, with $10
\lesssim M_{\star} \lesssim 140\,\msun$; and (4) Type II Supernova model 2,
with $30 \lesssim M_{\star} \lesssim 140\,\msun$. Model (2) is introduced to
see how sensitive the metal yield of PISN model
depends on various mass limits. Model (4) is motivated by the
suggestion of \citet{tan04} that the first stars should have masses between
$30-100\,\msun$. To avoid the degeneracy between the IMF
index and mass limits we adopted a consistent IMF index of
$\Gamma=1.5$ among all the four models.

For each IMF we compute the metal yield.
As pointed out by \citet{bro03} metal cooling
at low temperature ($T\le 100$K)
is dominated by fine-structure line cooling of \ion{C}{2} and
\ion{O}{1}, we concentrate on the metal yield of C and O.
The yield can be calculated through
\begin{equation}
y_i = \int_{m_1}^{m_2} \Psi(M) Y_i(M) dM,
\end{equation}
where the subscript $i$ represents C or O, and $Y_i(M)$ is defined as the
ratio of the mass fraction of the element $i$ in solar units, in this way
$Y_i(M)$ is one half of the production factors defined in \citet{heg02}.
For the PISN type IMF we use the production factors of \citet{heg02}. For
SNII type IMF model 1 and 2 we adopt the production factors calculated from
\citet{woo95} model Z12A, Z15A, ... and Z30B, Z35B, .... series,
respectively.

\begin{figure}
\begin{center}
\resizebox{3.5in}{!}{\includegraphics{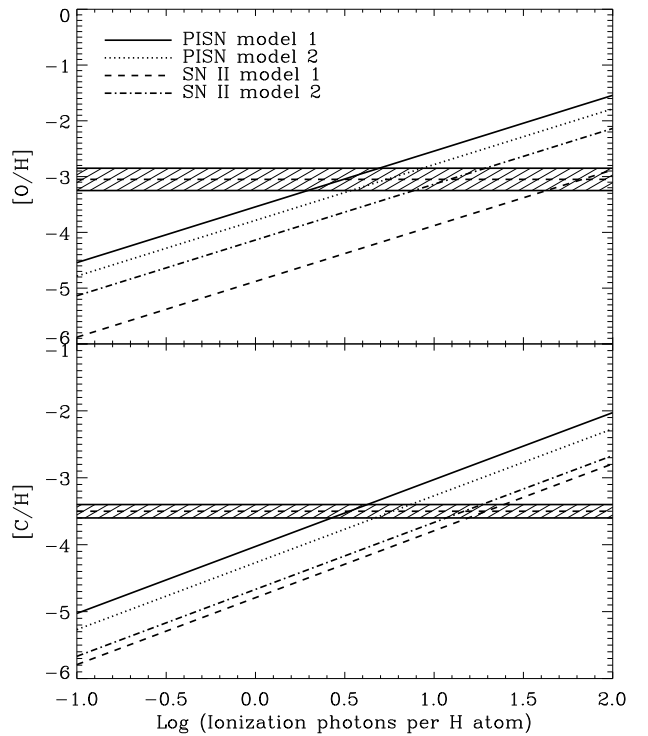}}
\caption{Oxygen (top panel) and Carbon (bottom panel) metallicities as a
function of the ionizing photons per H atom. The solid and dotted lines
represent PISN model 1 and 2, dashed and dot-dashed lines are for SNII model
1 and 2, respectively. The horizontal shadowed areas show the critical
transition metallicities and their errors from \citet{bro03}.}
\label{fig:metal}
\vskip7pt
\end{center}
\end{figure}

With metal yield $y_i$ and ionizing photon production rate $<N_{ph}>$, we
can now compute the element metallicity $Z^i_{IGM}$ as a function of
ionizing photons per hydrogen atom through Eq.~\ref{eq:ZIGM}. The results
are shown in Figure~\ref{fig:metal}. The top and bottom panels show the
oxygen and carbon metallicities, respectively. The metal
abundance in this definition is relative to the solar value, i.e., $\rm
[O/H] = \log_{10} (n_{O}/n_H) - \log_{10}
(n_{O}/n_H)_{\odot}$.  The solid and dotted lines
represent PISN model 1 and 2, dashed and dot-dashed lines are for SNII model
1 and 2, respectively. The horizontal shadowed areas show the critical
transition metallicities $Z_{crit}$ and their errors from \citet{bro03},
$\rm [C/H]_{crit} \simeq -3.5\pm0.1$ and $\rm [O/H]_{crit} \simeq
-3.05\pm0.2$, respectively.  The free-fall time scale for gas clump is
$t_{ff}
\approx 5\times 10^5 (n_H/10^4\,\rm cm^{-3})^{-1/2}\, yrs$, and the critical
transition metallicities are obtained by equating the cooling time at
temperature $T=100$K (below which metal cooling dominates) to $t_{ff}$ for a
gas density of $n_H=10^4$cm$^{-3}$
(\citealp{abe00,bro02}).

Immediately, we see that the transition from Pop III to Pop II occurs much
earlier in Model 1 and 2 than in Models 3 and 4.  We see that about
$1.0-3.1$ ionizing photons per H atom are produced in PISN Model 1, $3-6$
photons in PISN model 2, whereas
$8.6-14$ ionizing photons per H atom is produced in SN II Model 1 and 2. Not
all ionizing photons can escape into the IGM, nor do the
metals. Metal enrichment of the IGM may be expected to be
inhomogeneous, which perhaps would yield a higher effective metal
enrichment in star formation regions if galaxy formation is
biased. The fact supernova explosions in the PISN (Model 1) may be more
energetic than in the other two models may suggest that perhaps more metals
could be transported to the IGM in the former case.  It thus seems
improbable that PISN Pop III stars will be able to ionize the universe
alone, unless ionizing photon escape fraction is
$\ge 20\%$ and/or metal enrichment is very inefficient.

\section{Conclusions}

Based on the consideration of cooling due to C and O at low
temperatures ($T<100$K), we note that it may be expected that the
transition from Pop III to Pop II stars should occur when $\sim 1$
ionizing photon per H atom is produced, if the IMF is entirely VMS.  A
factor of about $10$ more ionizing photons would be produced in the
case with normal top-heavy IMF (e.g., $M\sim 10-100\,\msun$), when
such a transition occurs.  Thus, it will be much more difficult, if
not impossible, to achieve an earlier ($z\gg 10$) reionization in the
former case.  Therefore, if future WMAP data firm up the Thomson
optical depth to $\tau \ge 0.11-0.14$, one may be forced to adopt a
more conventional top-heavy IMF.  If minihalos were largely
responsible for producing ionizing photons at high redshift, then an
IMF with VMS may be still viable, as the ionizing photon escape
fraction from minihalos may well exceed $50\%$, shown by the recent
radiation hydrodynamic simulations (\citealp{wha04,kit04}).

We assume that metals can efficiently get mixed with the general IGM
and adopted a simplified picture that metals produced by Pop III stars
are uniformly distributed in the IGM.  In the PISN case \citet{brom03}
found for a supernova explosion with an energy of $\sim 10^{53}$ ergs,
at least 90\% of metals can be effectively transported into
surrounding IGM.  \citet{yos04} discussed a simplified VMS IMF (all
Pop III stars have mass of 200 $M_{\odot}$ and explode as PISN): they
trace the star formation history and calculate the mean number of
ionization photons per atom and the mean metalicity of the IGM.  Their
results are consistent with results from our PISN model 1 and 2.  The
derived $\tau_e$ falls short of $\sl WMAP$ result when constrained by
a critical metalicity of $[Z/H]=-3.3$, this is consistent with our
conclusion that a VMS IMF may not be favored by recent $\sl WMAP$
result.  However, metal mixing is most likely inhomogeneous, that
could affect the transition from Pop III to Pop II stars (see, e.g.,
\citealp{sca03,mac03,yos04}).  The exact effect is, however, unclear,
when coupled with spatially non-uniform galaxy and star formation.

We emphasize that a conventional IMF may not be the only way to
alleviate the conflict between the metal yield of the Pop III stars
and the number of ionization photons they produced. \citet{ric04}
argued that most Pop III stars should collapse into black holes to
avoid producing too much metals with added benefits of additional
accretion produced energy.

\smallskip
\acknowledgements{We thanks the anonymous referee for many
  helpful comments. We gratefully acknowledge financial support by
grants AST-0206299, AST-0407176 and NAG5-13381.
T.~Fang was supported by the NASA through
{\sl Chandra} Postdoctoral Fellowship Award Number PF3-40030 issued by the
{\sl Chandra} X-ray Observatory Center, which is operated by the
Smithsonian Astrophysical Observatory for and on behalf of the NASA under
contract NAS8-39073.}

\end{document}